\documentstyle[prd,aps]{revtex}

\newcommand{\bea}{\begin{eqnarray}}
\newcommand{\eea}{\end{eqnarray}}

\begin{document}
\draft
\twocolumn[\hsize\textwidth\columnwidth\hsize\csname
@twocolumnfalse\endcsname

\title{Conserved quantities in the perturbed Friedmann world model}
\author{Jai-chan Hwang}
\address{Department of Astronomy and Atmospheric Sciences,
         Kyungpook National University, Taegu, Korea}

\date{\today}
\maketitle

\begin{abstract}

The evolutions of linear structures in a spatially homogeneous and isotropic
world model are characterized by some conserved quantities:
the amplitude of gravitational wave is conserved in the super-horizon scale,
the perturbed three-space curvature in the comoving gauge is conserved 
in the super-sound-horizon scale, and the angular momentum of rotational
perturbation is generally conserved.

\end{abstract}

\noindent

\vskip2pc]

{\it 1. Summary:}
We consider a spatially homogeneous and isotropic world model 
with the most general, spacetime dependent, three (scalar, vector, and tensor) 
types of linear structures.  
As the gravity sector we consider both Einstein gravity with a hydrodynamic 
fluid or a scalar field, and a class of generalized versions of gravity 
which couples the scalar field with the scalar curvature.
The three types of structures decouple from each other due to the symmetry
in the background world model and the linearity of the structures we are
assuming.
We identify the generally conserved quantities existing in all three types 
of perturbations, ignoring the imperfect fluid source terms:

(A) 
In a near flat background the scalar-type (related to the density condensation) 
structure is characterized by a conserved quantity 
in the large-scale which is the perturbed three-space curvature 
in the comoving (or the uniform-field) gauge.
This quantity is conserved in the super-sound-horizon scale, and
thus is conserved effectively in all scales in the matter dominated era.

(B) 
The vector-type (rotation) structure is characterized by the 
angular momentum conservation.

(C) 
In a near flat (vanishing spatial curvature) background the amplitude 
of the tensor-type (gravitational wave) structure is conserved in 
the super-horizon scale.

In a similar sense, the sign of three-space curvature in the underlying 
world model, $K$, can be also regarded as a conserved quantity.

In the fluid and the scalar field dominated eras the conserved 
properties remain valid independently of changing (i.e., time-varying) 
equation of state $p(\mu)$ and changing field potential $V (\phi)$, 
respectively.
These conservation properties also apply in a class of
generalized versions of gravity theories including the Brans-Dicke theory,
the non-minimally coupled scalar field, the scalar-curvature-square gravity,
the low-energy effective action of the string theory, etc.
As long as the conditions are met the conservation properties remain valid
independently of changing gravity theories.
The conservation properties in (B,C) are valid in the multi-component 
situations, whereas (A) applies to single-component situations.

In the following we will present the above mentioned results by
concretely showing the equations describing the results.
Since this is a summary of our previous works, 
the details about derivations and explanations are referred to 
the original works in the literature.

\vskip .5cm
{\it 2. Gravity theories:}
We consider gravity theories belonging to the following action
\bea
   & & S = \int d^4 x \sqrt{-g} \Big[ {1 \over 2} f (\phi, R)
       - {1\over 2} \omega (\phi) \phi^{;a} \phi_{,a} - V(\phi)
   \nonumber \\
   & & \qquad \qquad \qquad \qquad
       + L_m \Big],
   \label{action}
\eea
where $\phi$ and $R$ are the scalar field and the scalar curvature,
respectively.
$L_m$ is the hydrodynamic part of the Lagrangian with the hydrodynamic 
energy-momentum tensor $T_{ab}$ defined as $\delta ( \sqrt{-g} L_m )
\equiv {1 \over 2} \sqrt{-g} T^{ab} \delta g_{ab}$.

\vskip .5cm
{\it 3. Cosmological perturbations:}
We consider spacetime dependent perturbations in the background
Friedmann world model
\bea
   d s^2 
   &=& - \left( 1 + 2 \alpha \right) d t^2
       - a \left( \beta_{,\alpha} +B_\alpha \right) d t d x^\alpha
   \nonumber \\
   & & + a^2 \Big[ g_{\alpha\beta}^{(3)} \left( 1 + 2 \varphi \right)
       + 2 \gamma_{|\alpha\beta} 
   \nonumber \\
   & & \qquad \quad
       + 2 C_{(\alpha|\beta)} + 2 C_{\alpha\beta}
       \Big] d x^\alpha d x^\beta.
\eea
$\alpha$, $\beta$, $\gamma$, and $\varphi$ 
indicate the scalar-type structure with four degrees of freedom.
The transverse $B_\alpha$ and $C_\alpha$
indicate the vector-type structure with four degrees of freedom.
The transverse-tracefree $C_{\alpha\beta}$
indicates the tensor-type structure with two degrees of freedom.
The three types of structures are related to the density condensation,
the rotation, and the gravitational wave, respectively.
Since these three types of structures evolve independently, we will handle
them separately.

We also consider the general perturbations in the hydrodynamic 
energy-momentum tensor and the scalar field:
\bea
   & & T_{ab} ({\bf x}, t) = \bar T_{ab} (t) + \delta T_{ab} ({\bf x}, t), 
   \nonumber \\
   & & \phi ({\bf x}, t) = \bar \phi (t) + \delta \phi ({\bf x}, t).
\eea

\vskip .5cm
{\it 4. Gauge issue:}
We have ten degrees of freedom in the metric perturbation.
Two (one temporal and one spatial) degrees of freedom in the scalar-type 
perturbation, and two (both spatial) degrees of freedom in the vector-type
perturbation are due to the spacetime coordinate (gauge) transformation.
In order to handle these fictitious degrees of freedom we have the
right to impose four conditions on the respective perturbed metric
or energy-momentum content.
Due to the spatial symmetry (homogeneity) the three spatial degrees of
freedom can be uniquely fixed, thus making the remaining variables
spatially gauge-invariant \cite{Bardeen-1988};
for example, $\chi \equiv a (\beta + a \dot \gamma)$ is a
spatially gauge-invariant combination.
After removing the spatial gauge degrees of freedom, only the scalar-type
perturbed variables are affected by the temporal gauge condition.
When we fix the temporal gauge condition, we have {\it several} 
meaningful choices.
Since usually we do not know which gauge condition will turn out to be
most useful for the problem {\it a priori}, it is advantageous to start 
from equations without fixing the temporal gauge condition, 
thus in a gauge-ready form, \cite{Bardeen-1988,PRW}.

Using the gauge-ready approach we have investigated the cosmological 
perturbation based on the hydrodynamic fluid, scalar field, and 
a class of generalized gravity, \cite{PRW,Ideal,MDE,MSF,GGT-HN}.
{}From these studies we find the best conserved quantity is the
perturbed three-space curvature $\varphi$ in the comoving gauge
($v \equiv 0$) or in the uniform-field gauge ($\delta \phi \equiv 0$);
in the generalized gravity theory the uniform-field gauge is the better
gauge condition, and the uniform-field gauge coincides with the 
comoving gauge in the minimally coupled scalar field.
Both gauge conditions completely fix the gauge transformation and
each variable evaluated in the gauge is the same as the
corresponding unique gauge invariant combination of the variable
and a variable used in the gauge condition.
Since we have many different gauge conditions we proposed to write
the gauge-invariant combinations in the following way, \cite{PRW}:
\bea
   \varphi_v \equiv \varphi - {aH \over k} v, \quad
       \varphi_{\delta \phi} \equiv \varphi - {H \over \dot \phi} \delta \phi
       \equiv - {H \over \dot \phi} \delta \phi_\varphi,
   \label{varphi-delta-phi}
\eea
where $k$ is a comoving wavenumber and $H \equiv \dot a/a$. 
$\varphi_v$ is the same as $\varphi$ in the comoving gauge condition
which takes $v \equiv 0$, etc; $v$ is a velocity related perturbed variable, 
\cite{Newtonian}.

\vskip .5cm
{\it 5. Evolution equations:}
The equations for background are:
\bea
   & & H^2 = {1 \over 3F} \left[ \mu
       + {1 \over 2} \left( \omega \dot \phi^2
       - f + RF + 2V \right) - 3 H \dot F \right]
       - {K \over a^2},
   \nonumber \\
   \label{BG-1} \\
   & & \ddot \phi + 3 H \dot \phi
       + {1 \over 2 \omega} \left( \omega_{,\phi} \dot \phi^2
       - f_{,\phi} + 2 V_{,\phi} \right) = 0,
   \label{BG-2} \\
   & & \dot \mu + 3 H \left( \mu + p \right) = 0,
   \label{BG-3}
\eea
where $F \equiv \partial f / (\partial R)$.
$K$ is the sign of the background spatial curvature. 
It can be considered as an integration constant, and thus is a conserved 
quantity for the given world model.

In the following we summarize the evolution equations for three types
of perturbations.
We consider {\it near flat} background, thus neglect $K$ term.

(A) 
In handling the scalar-type structure the proper choice of the gauge
condition simplifies the analyses and the resulting equations.
A gauge invariant combination of the perturbed curvature variable
based on the comoving (or uniform-field) gauge is found to be the best 
conserved quantity under various changes.
The equations describing the evolutions of the hydrodynamic \cite{Newtonian}, 
the minimally coupled scalar field \cite{H-QFT}, and the generalized gravity 
theories \cite{GGT}, respecively, are the following:
\bea
    & & {c_s^2 H^2 \over a^3 ( \mu + p)} \left[ {a^3 ( \mu + p) \over c_s^2 H^2}
        \dot \varphi_v \right]^\cdot
        - c_s^2 {\Delta \over a^2} \varphi_v = {\rm stresses},
    \label{varphi-eq1} \\
    & & {H^2 \over a^3 \dot \phi^2} \left( {a^3 \dot \phi^2 \over H^2}
        \dot \varphi_{\delta \phi} \right)^\cdot
        - {\Delta \over a^2} \varphi_{\delta \phi} = 0,
    \label{varphi-eq2} \\
    & & {\left(H + {\dot F \over 2F} \right)^2 \over a^3 
        \left( \omega \dot \phi^2 +{3 \dot F^2 \over 2 F} \right)} 
        \left[ {a^3 \left( \omega \dot \phi^2 +{3 \dot F^2 \over 2 F} \right)
        \over \left(H + {\dot F \over 2F} \right)^2 }
        \dot \varphi_{\delta \phi} \right]^\cdot
        - {\Delta \over a^2} \varphi_{\delta \phi} = 0,
    \nonumber \\
    \label{varphi-eq3} 
\eea
where $\Delta$ is a Laplacian operator based on $g_{\alpha\beta}^{(3)}$
and $c_s^2 \equiv \dot p/ \dot \mu$.
{}For a pressureless ideal fluid, instead of Eq. (\ref{varphi-eq1})
we have $\dot \varphi_v = 0$.

(B)
We introduce $B_\alpha \equiv b Y^{(v)}_\alpha$ and
$C_\alpha \equiv c Y^{(v)}_\alpha$ where $Y^{(v)}_\alpha$ is a 
(transverse) vector harmonic function, \cite{Bardeen,Rab-rot}.
$v_\omega \equiv v^{(v)} - b$ is a gauge invariant combination related to 
the amplitude of hydrodynamic vorticity $\omega$ as $v_\omega \propto a \omega$, 
\cite{Rab-rot}.
The rotational structure is described by
\bea
   \left[ a^4 \left( \mu + p \right) v_\omega \right]^\cdot
       = {\rm stress}.
   \label{rot-eq}
\eea
Thus, neither the generalized nature of the gravity theory nor the presence of
scalar field affects the rotational perturbation in the hydrodynamic part.
We note that Eq. (\ref{rot-eq}) is valid even in a generalized
gravity with the Ricci-curvature-square term in the action, \cite{Rab-rot}.

(C)
The gravitational wave is described by \cite{PRW,GGT-GW}
\bea
   {1 \over a^3 F} \left( a^3 F \dot C^\alpha_\beta \right)^\cdot
       - {\Delta \over a^2} C^\alpha_\beta = {\rm stress}.
   \label{GW-eq}
\eea

\vskip .5cm
{\it 6. Solutions:}
Equations (\ref{varphi-eq1}-\ref{GW-eq}) immediately lead to general 
solutions in certain limiting situations.

(A)
{}From Eqs. (\ref{varphi-eq1}-\ref{varphi-eq3}) we have the large-scale 
solution, for the hydrodynamic (ignoring the stresses),
the scalar field, and the generalized gravity: 
\bea
   & & \varphi_v 
       = C ({\bf x}) - \tilde D ({\bf x}) 
       \int^t_0 {c_s^2 H^2 \over a^3 ( \mu + p ) } dt,
   \label{varphi-sol1} \\
   & & \varphi_{\delta \phi} 
       = C ({\bf x}) - D ({\bf x}) \int^t_0 {H^2 \over a^3 \dot \phi^2 } dt,
   \label{varphi-sol2} \\
   & & \varphi_{\delta \phi} 
       = C ({\bf x}) - D ({\bf x}) \int^t_0 
       {\left(H + {\dot F \over 2F} \right)^2 \over a^3
       \left( \omega \dot \phi^2 +{3 \dot F^2 \over 2 F} \right)} dt,
   \label{varphi-sol3}
\eea
where $C$ and $D$ (or $\tilde D$) are integration constants indicating 
coefficients of relatively growing and decaying solutions, respectively.
Compared with the solutions in the other gauge conditions,
the decaying modes in these solutions are {\it higher order} in the
large-scale expansion, \cite{GGT-HN,Newtonian}.
Thus, ignoring the transient (and also subdominating in the large-scale)
modes we have the conserved quantity 
\bea
   \varphi_{\delta \phi} = C ({\bf x}) = \varphi_v.
   \label{C}
\eea
{}From Eqs. (\ref{varphi-eq1}-\ref{varphi-eq3}) we notice that 
Eq. (\ref{varphi-sol1}) is valid in the super-sound horizon scale,
whereas Eqs. (\ref{varphi-sol2},\ref{varphi-sol3}) are valid 
in the super-horizon scale.

In the large-scale limit $\varphi$ in many different gauge conditions
shows the conserved behavior: in the case of an ideal fluid see 
Eqs. (41,73) in \cite{Ideal}, and Eq. (34,35) in \cite{MDE}; 
in the case of the scalar field see Eqs. (92) in \cite{MSF};
and in the case of the generalized gravity see Sec. VI in \cite{GGT-HN}.
The often discussed conserved variable $\zeta$ introduced in \cite{BST}
is $\varphi_{\delta}$ which is $\varphi$ in the uniform-density gauge.
In \cite{Newtonian} we made arguments why we regard $\varphi_v$ as the best
conserved quantity.
Conservation properties are further discussed in \cite{Conserv}.

(B)
{}For vanishing anisotropic stress, Eq. (\ref{rot-eq}) for the rotation mode
has a solution \cite{Lifshitz,Rab-rot}
\bea
   a^3 (\mu + p) \cdot a \cdot v_\omega \sim L ({\bf x}).
   \label{rot-sol}
\eea
Thus, the rotation mode of the hydrodynamic part is characterized by
a conservation of the angular momentum $L$.

(C)
In the large-scale limit (super-horizon scale), 
ignoring the anisotropic stress, from Eq. (\ref{GW-eq}) we have a general 
solution for the gravitational wave \cite{GW-sol}
\bea
   C^\alpha_\beta ({\bf x}, t)
       = c^\alpha_\beta ({\bf x})
       - d^\alpha_\beta ({\bf x}) \int^t_0 {1 \over a^3 F} dt,
   \label{GW-sol}
\eea
where $c^\alpha_\beta$ and $d^\alpha_\beta$ are integration constants 
indicating coefficients of relatively growing and decaying solutions, 
respectively.
Ignoring the transient mode, the amplitude of gravitational wave
in the super-horizon scale is temporally conserved.

\vskip .5cm
{\it 7. Unified forms:}
The equations and the large-scale solutions for the scalar- and tensor-type
structures can be written in unified forms as:
\bea
   & & {1 \over a^3 Q} (a^3 Q \dot \Phi)^\cdot - c_A^2 {\Delta \over a^2} \Phi 
       = 0,
   \\
   & & \Phi = C ({\bf x}) - D ({\bf x}) \int_0^t (a^3 Q)^{-1} dt,
\eea
where, for the scalar-type fluid, the generalized gravity, and 
the tensor-type structures, respectively, we have:
\bea
   & & \Phi = \varphi_v, \quad \; Q = { \mu + p \over c_s^2 H^2 }, 
       \qquad \qquad \quad \; c_A^2 \rightarrow c_s^2,
   \\
   & & \Phi = \varphi_{\delta \phi}, \quad
       Q = { \omega \dot \phi^2 + 3 \dot F^2 / 2 F
       \over \left( H + \dot F / 2 F \right)^2 },
       \quad c_A^2 \rightarrow 1,
   \\
   & & \Phi = C^\alpha_\beta, \quad \; Q = F,
       \qquad \qquad \qquad \quad c_A^2 \rightarrow 1.
\eea

\vskip .5cm
{\it 8. Applications:}
We have shown conserved quantities in all three types of structures
in the Friedmann world model.
Even in the super-horizon scale, not every quantity is conserved,
and rather there exist some special (gauge-invariant)
variables which are conserved independently of changing equation of state
$p(\mu)$, changing scalar field potential $V(\phi)$, 
and changing gravity theories $f(\phi,R)$.
Using the conserved quantities, as long as the linearity assumption is valid,
we can easily trace the evolution of structures from the recent era to 
the early stage.

Let us consider a scenario where a generalized gravity (or Einstein gravity
with a scalar field) dominates the early evolution stage of our observable 
patch of the universe, and
at some point Einstein gravity takes over the dominance till the present era.
If we further assume that the generalized gravity era provides an
accelerated expansion (inflation) stage, the observationally relevant
scales may exit the horizon to become super-horizon scales during the era.
Under some conditions we can derive the generated quantum
fluctuations based on the vacuum expectation value,
and as the scale becomes the super-horizon it can be interpreted 
as the classical fluctuations based on the spatial average.
As long as the relevant scales remain in the super-horizon stage during the
transit epoch of the gravities (or changing potential, or changing
equation of state) there exist {\it conserved quantities}:
$C^\alpha_\beta$ and $\varphi_v$ (or $\varphi_{\delta \phi}$) 
are the conserved quantities.

Now, let us explain more concretely how the conserved quantities provide
connections between the hydrodynamic perturbations in Newtonian regime 
and the quantum fluctuations in the early universe.
We consider the scalar-type structures; for tensor-type structures, see
\cite{GGT-GW}. 
The quantum generation process is most easily handled using 
a perturbed scalar field ($\delta \phi$) equation
in the uniform-curvature gauge ($\varphi \equiv 0$), i.e., using 
$\delta \phi_\varphi$.
The analytic forms of power-spectrum of $\delta \phi_\varphi$ are available 
(in general scales) in variety of inflation stages based on the scalar 
field and the generalized gravity, \cite{GGT-application}.
Using Eq. (\ref{varphi-delta-phi}) the power-spectrum of $\delta \phi_\varphi$
is directly related to the power-spectrum of $\varphi_{\delta \phi}$, and the
latter quantity is conserved during a super-horizon scale evolution
which may be the case for the observationally relevant large-scale structures 
in post inflationary era.
As long as the scale remains in the super-horizon it is conserved 
independently of the reheating process, possible gravity change
(e.g., from the generalized gravity to Einstein one), and the change
from the scalar field dominated to fluid dominated stages;
i.e., $\varphi_{\delta \phi} = C ({\bf x}) = \varphi_v$.
Later on, in the hydrodynamic stage, from $\varphi_v$ we can derive 
the rest of perturbation variables; afterall, since we are considering
the linear perturbation, a variable is a linear combination of the other
variables.
It is known that $\varphi_\chi$ ($\varphi$ in the zero-shear gauge
which takes $\chi \equiv 0$),
$\delta_v$ ($\delta \equiv \delta \mu/\mu$ in the comoving gauge),
and $v_\chi$ ($v$ in the zero-shear gauge) closely correspond to 
the Newtonian potential fluctuation, density contrast, and velocity 
fluctuations, respectively, \cite{Bardeen,Newtonian}.
We have \cite{Newtonian}:
\bea
   & & \varphi_\chi = C \left( 1 - {H \over a} \int_0^t a dt \right)
       + {H \over a} d, 
   \\
   & & \delta_v = {2 k^2 \over \mu a^2} \varphi_\chi,
   \\
   & & v_\chi = {k \over a^2} \left( - C \int_0^t a dt + d \right),
\eea
which are valid for general $p(\mu)$, but assuming $K = 0$ and 
vanishing stresses.
The observed anisotropy of the cosmic microwave background radiation 
in the large angular scale is also related to $C$ at the last scattering epoch 
as ${\delta T/T} = - {1 \over 5} C$, 
where we assumed a matter dominated era and ignored the decaying mode.
As mentioned, $\tilde D$ terms in Eq. (\ref{varphi-sol1}) is $(k/aH)^2$ 
higher order compared with $d ({\bf x})$ term in these solutions.

Thus, in order to determine the Newtonian quantities ($\varphi_\chi$,
$\delta_v$, and $v_\chi$ which provide the initial conditions for later 
nonlinear evolution stage) what we need is the spatial structures encoded in 
$C({\bf x})$.
{}From Eqs. (\ref{varphi-delta-phi},\ref{C}) the power-spectrum of
spatial distribution of $C$, ${\cal P_C}$, can be directly related to
the power-spectrum of $\delta \phi_\varphi$ in the inflationary stage as
\bea
   {\cal P}_C^{1/2} = {\cal P}_{\varphi_v}^{1/2}  
   = {\cal P}_{\varphi_{\delta \phi}}^{1/2} 
   = {H \over |\dot \phi|} {\cal P}_{\delta \phi_\varphi}^{1/2}.
\eea
The evaluation of quantum fluctuations of $\delta \phi_{\varphi}$
based on the vacuum expectation value leads to ${\cal P}_{\delta \phi_\varphi}$;
for example, in near-exponential inflation based on scalar field,
in the large-scale and in a simplest vacuum choice, we have
${\cal P}^{1/2}_{\delta \phi_\varphi} = H/2\pi$, \cite{QFT,GGT-application}.
In this paradigm of large-scale structures generated from quantum fluctuations,
we can probe the physics in inflation stage using the observed large-scale 
structures and the anisotropy in the cosmic microwave background radiation. 
Recent endeavors for reconstructing and constraining the inflation physics 
from observation can be found in \cite{reconstruction}.

\vskip .5cm
Note added: After publication of this work we found a new quantity for 
hydrodynamic scalar-type perturbation which is conserved considering 
general $K$, $\Lambda$, and time-varying $p(\mu)$:
see \cite{Hydro}

\vskip .5cm
We thank Dr. H. Noh for useful discussions.
This work was supported by the Korea Science and Engineering Foundation, 
Grant No. 95-0702-04-01-3 and through the SRC program of SNU-CTP.



\begin{thebibliography}{99}
\bibitem{Bardeen-1988}
         J. M. Bardeen in {\it Particle Physics and Cosmology}, edited by
               L. Fang and A. Zee
               (London, Gordon and Breach, 1988), 1.
\bibitem{PRW}
         J. Hwang, Astrophys. J. {\bf 375}, 443 (1991).
\bibitem{Ideal}
         J. Hwang, Astrophys. J. {\bf 415}, 486 (1993).
\bibitem{MDE}
         J. Hwang, Astrophys. J. {\bf 427}, 533 (1994).
\bibitem{MSF}
         J. Hwang, Astrophys. J. {\bf 427}, 542 (1994).
\bibitem{GGT-HN}
         J. Hwang and H. Noh, Phys. Rev. D {\bf 54}, 1460 (1996).
\bibitem{Newtonian}
         J. Hwang and H. Noh, preprint astro-ph/9907063.
\bibitem{H-QFT}
         J. Hwang, Phys. Rev. D {\bf 48}, 3544 (1993).
\bibitem{GGT}
         J. Hwang, Phys. Rev. D {\bf 53}, 762 (1996).
\bibitem{Bardeen}
         J. M. Bardeen, Phys. Rev. D {\bf 22}, 1882 (1980).
\bibitem{Rab-rot}
         J. Hwang and H. Noh, Phys. Rev. D {\bf 57}, 2617 (1998).
\bibitem{GGT-GW}
         J. Hwang and H. Noh, Class. Quant. Grav. {\bf 15}, 1401 (1998).
\bibitem{BST}
         J. M. Bardeen, P. J. Steinhardt and M. S. Turner,
               Phys. Rev. D {\bf 28}, 679 (1983).
\bibitem{Conserv}
         D. H. Lyth, Phys. Rev. D {\bf 31}, 1792 (1985);
         D. H. Lyth and M. Mukherjee, Phys. Rev. D {\bf 38}, 485 (1988);
         J. Hwang, Astrophys. J. {\bf 380}, 307 (1991);
         J. Hwang and J. J. Hyun, Astrophys. J. {\bf 420}, 512 (1994);
         W. Zimdahl, Class. Quant. Grav. {\bf 14}, 2563 (1997).
\bibitem{Lifshitz}
         E. M. Lifshitz, J. Phys. (USSR) {\bf 10}, 116 (1946);
         E. M. Lifshitz and I. M. Khalatnikov, Adv. Phys. {\bf 12}, 185 (1963);
         L. D. Landau and E. M. Lifshitz, {\it The Classical Theory of Fields},
               4th Edition (Pergamon, Oxford, 1975), Sec. 115.
\bibitem{GW-sol}
         A. A. Starobinsky, JETP Lett. {\bf 30}, 682 (1979);
         J. Hwang, Class. Quant. Grav. {\bf 8}, 195 (1991).
\bibitem{GGT-application}
         J. Hwang, Class. Quant. Grav. {\bf 14}, 3327 (1997);
         J. Hwang and H. Noh, Class. Quant. Grav. {\bf 15}, 1387 (1998).
\bibitem{QFT}
         D. H. Lyth and E. D. Stewart, Phys. Lett. B {\bf 274}, 168 (1992);
         J. Hwang, Phys. Rev. D {\bf 48}, 3544 (1993).
\bibitem{reconstruction}
         A. R. Liddle and D. H. Lyth, Phys. Rep. {\bf 231}, 1 (1993);
         J. E. Lidsey, etal., Rev. Mod. Phys. {\bf 69}, 373 (1997),
         and references therein.
\bibitem{Hydro}
         J. Hwang, preprint, astro-ph/9907080.
\end{thebibliography}
\end{document}